\documentclass[times,twocolumn,final]{elsarticle}
\usepackage{cmpb}
\usepackage{framed,multirow}
\usepackage{booktabs,array,colortbl}
\usepackage{amsmath,amssymb,amsfonts}
\usepackage{textcomp}
\usepackage{url}
\usepackage{graphicx}
\usepackage{caption}
\usepackage{rotating}
\usepackage{xcolor}
\usepackage{atbegshi}

\definecolor{BestGreen}{HTML}{E6F4EA}
\newcommand{\best}[1]{\cellcolor{BestGreen}\textbf{#1}}

\journal{Results in Engineering}
\begin{document}

\begin{frontmatter}

\title{NL-MambaXCT: Self-Supervised Nested-Learning Mamba for Nomex Honeycomb X-ray CT Defect Classification}

\author[aurak,strata,mbzuai]{Ghaleb Aldoboni\fnref{equal}}
\ead{ghaleb.aldoboni@aurak.ac.ae}
\ead{ghaleb.aldoboni@mbzuai.ac.ae}

\author[aurak]{Lobna Nassar\fnref{equal}}
\ead{lobna.nassar@aurak.ac.ae}

\author[mbzuai,uwaterloo]{Fakhri Karray\corref{cor1}}
\ead{fakhri.karray@mbzuai.ac.ae}
\ead{karray@uwaterloo.ca}

\author[strata]{Reem Alshamsi}
\ead{rsalshamsi@strata.ae}

\fntext[equal]{These authors contributed equally to this work.}

\cortext[cor1]{Corresponding author}

\affiliation[aurak]{
  organization={Department of Computer Science and Engineering, American University of Ras Al Khaimah (AURAK)},
  city={Ras Al Khaimah},
  country={United Arab Emirates}
}

\affiliation[strata]{
  organization={Strata Manufacturing},
  city={Abu Dhabi},
  country={United Arab Emirates}
}

\affiliation[mbzuai]{
  organization={Department of Machine Learning, Mohamed bin Zayed University of Artificial Intelligence (MBZUAI)},
  city={Abu Dhabi},
  country={United Arab Emirates}
}

\affiliation[uwaterloo]{
  organization={Centre for Pattern Analysis and Machine Intelligence, Department of Electrical and Computer Engineering, University of Waterloo},
  city={Waterloo},
  state={Ontario},
  postcode={N2L 3G1},
  country={Canada}
}


\begin{abstract}
X-ray computed tomography (XCT) is widely used for non-destructive testing of Nomex honeycomb structures in aerospace manufacturing, yet industrial inspection workflows still rely heavily on manual interpretation and supervised models trained on small, noisy datasets. This work introduces NL-MambaXCT, a Mamba-based framework that combines self-supervised masked image modelling with a Nested Learning (NL) formulation to enable automated, label-efficient defect classification from production XCT data. The NL-MambaXCT constitutes the first fully automated deep-learning pipeline for XCT inspection of Nomex honeycomb structures and the first application of Nested Learning to image-level classification in non-destructive testing. The backbone is a four-stage 2D encoder that uses RegNet convolutional blocks in the early stages and Mamba-based sequence mixing with attention in the deeper stages, and is pretrained via masked image modelling on 19{,}961 unlabeled industrial XCT slices before being fine-tuned on a curated dataset of 2{,}000 relabeled Nomex XCT slices split by production order. NL is instantiated through two-timescale parameter dynamics, in which key projections maintain slow exponential-moving-average traces alongside fast weights. A deep-momentum optimizer introduces an additional slow timescale at the level of parameter updates, stabilizing training under limited and imbalanced labels. On the test set, the MIM-pretrained NL-MambaXCT model achieves 96.8\% accuracy and a macro F1-score of 96.7\%, outperforming strong CNN, attention, and single-timescale Mamba baselines by 4.1--10.2 percentage points in accuracy and retaining a clear advantage when the amount of labeled data is reduced. In a sequential production-batch evaluation, it obtains 96.64\% held-out macro F1 with only 0.08\% mean catastrophic forgetting. These results indicate that combining masked self-supervision on unlabeled XCT with NL-style fast/slow learning dynamics is a promising strategy for robust, industrial-grade defect classification in Nomex honeycomb XCT inspection.
\end{abstract}


\end{frontmatter}
\section{Introduction}
Honeycomb sandwich structure (HCSS) materials are used in the maritime and aerospace industries due to their favorable stiffness-weight ratio and durability \cite{Nunes2016Sandwiched}. However, these structures are vulnerable to defects introduced during manufacturing or service life, including deformation, displacement, split, foreign object damage (FOD) \cite{Han2022Numerical,Habermehl2012Multiple-frequency}. Such defects can severely compromise the structural integrity and lead to premature failure, necessitating a reliable inspection and classification of discontinuities that are essential for safe operation and cost-effective maintenance. Hence, a wide range of non-destructive testing (NDT) techniques is used to detect damage in HCSS, including radiography, X-ray computed tomography (XCT), shearography, thermography, electronic speckle pattern interferometry (ESPI) and ultrasound tests \cite{Choi2008Inspection}. Each modality offers a distinct trade-off between sensitivity, penetration depth, acquisition speed, and is specialized in detecting a specific set of deficits. In practice, the selection of the method depends on the morphology of the defect, the geometry of the components, and industrial constraints such as inspection throughput and accessibility \cite{Akatay2015The}. In industrial aerospace environments, XCT inspection of HCSS components is still predominantly performed by highly trained NDT experts, who manually inspect individual slices or volume renderings and cross-reference findings with production documentation. This process is labor-intensive, might be subjective, and difficult to scale for high-throughput production lines.
\subsection{Literature Review}
\subsubsection{Deep Learning for NDT and Dataset Limitations}
Recent advances in artificial intelligence have significantly affected NDT. Convolutional and recurrent neural networks have been successfully applied to thermography, ultrasound, and radiographic imaging for the automatic detection and classification of defects. For example, dynamic infrared scanning thermography combined with convolutional neural networks has achieved high defect detection accuracy with short training times \cite{Li2024}. Meanwhile, LSTM-based models have been utilized to classify honeycomb core defects from dynamic infrared sequences with high sensitivity \cite{hu2019lstm}. Other Deep-learning-based approaches have also been shown to meet industrial standards for weld inspection \cite{Naddaf2022Defect}, to recognize debonding and other defects in aerospace honeycomb structures using thermography \cite{gao2023convolution}, and to automate defect analysis in radiographic weld images \cite{Boaretto2017Automated}.

Despite these successes, the deployment of supervised deep learning in industrial NDT remains constrained by the availability and quality of labeled data. Accurate annotation of defect type and location requires expert knowledge and is time-consuming, and annotation noise can severely degrade model performance \cite{cui2024impact}. Surveys of deep learning in NDT and industrial inspection repeatedly point to data scarcity, class imbalance, and costly labeling workflows as the main obstacles to reliable deployment \cite{alzubaidi2023survey,ren2020survey,ahmed2023deep}. These limitations are particularly pronounced in XCT, where individual scans can contain hundreds of slices, yet only a small fraction may exhibit critical defects.

\subsubsection{Unsupervised and Self-Supervised Learning in NDT}
To mitigate dependence on large labeled datasets, unsupervised and self-supervised learning approaches have gained increasing attention in NDT, as summarized in Table~\ref{tab:ssl_ndt} \cite{tsai2021,chow2020,kim2024,huang2023,pang2024}. Autoencoder-based architectures, variational autoencoders, and clustering-based frameworks have been used to learn representations of normal material behavior from defect-free data, enabling anomaly detection without explicit defect labels; refer to Table~\ref{tab:unsupervised_summary}. 
\begin{table*}[h]
\centering
\caption{Representative applications of self-supervised learning in nondestructive testing (NDT)}
\label{tab:ssl_ndt}
\small
\setlength{\tabcolsep}{3pt}
\begin{tabular}{p{2.2cm} p{3.4cm} p{3.2cm} p{4.6cm} p{3.2cm}}
\hline
\textbf{Modality} & \textbf{Task} & \textbf{Role of SSL} & \textbf{Architecture} & \textbf{Citations} \\
\hline
Ultrasonic &
Flaw classification, near-surface defect detection &
Representation learning, anomaly detection &
DINO self-distillation SSL (commonly with a ViT backbone); domain-knowledge-informed SSL with synthetic defects and a de-anomaly network &
\cite{kim2024,jeon2024ultrasonic} \\

X-ray / Radiography &
Weld and casting defect detection &
Domain-specific pretraining, denoising &
SS-FDPNet self-supervised denoising network with frequency-band processing; SimSiam/SimMIM SSL pretraining using CNN and Swin Transformer backbones &
\cite{zhang2025,lol} \\

GPR &
Tunnel lining inspection &
Contrastive pretraining for detection backbone &
SA-DenseCL (self-attention DenseCL) pretraining; transfer to Mask R-CNN for detection/segmentation &
\cite{huang2023} \\

Hyperspectral &
Food adulteration detection &
Imbalanced classification with SSL pretraining &
Proto-DS: contrastive SSL pretraining + prototypical networks with Dice-based optimization for imbalance &
\cite{pang2024} \\
\hline
\end{tabular}
\end{table*}

\begin{table*}[h]
\centering
\caption{Summary of unsupervised reconstruction- and clustering-based defect detection frameworks}
\label{tab:unsupervised_summary}
\small
\setlength{\tabcolsep}{3pt}
\begin{tabular}{p{2.2cm} p{3.5cm} p{2.3cm} p{2.6cm} p{3.2cm} p{2.7cm}}
\hline
\textbf{Framework} & \textbf{Architecture} & \textbf{Training data} & \textbf{Anomaly cue} & \textbf{NDT/industrial examples} & \textbf{Citations} \\
\hline
CAE / Denoising AE &
Convolutional encoder--decoder; latent-center regularization; denoising AE with multi-feature fusion (e.g., FeatDAE); baseline-optimized AE (SHM) &
Defect-free only &
Reconstruction error or feature distance &
Surface inspection, concrete, guided waves &
\cite{tsai2021,chow2020,tunukovic2024,zhou2025,yang2025} \\

VAE / Diffusion / GAN &
VAE probabilistic latent model; diffusion-based detector; adversarial generative variants; VAE-based synthetic defect generation for augmentation &
Defect-free (plus few defects for augmentation) &
Reconstruction error, low likelihood, synthetic data &
UT, LUVT, semiconductor manufacturing &
\cite{chadha2021,tunukovic2024,shiferaw2024,ando2024,fan2023} \\

AE + Clustering &
Split-latent deep AE (discriminative vs.\ reconstructive) with k-means-style clustering; DBSCAN on reconstruction residuals; CNN head for defect class &
Defect-free (sometimes mixed) &
Cluster outliers in latent / error space &
Gear wheels, industrial time series, SHM &
\cite{chadha2021,klarak2024} \\
\hline
\end{tabular}
\end{table*}

Self-supervised pre-training on domain-specific X-ray data has recently been shown to enhance defect detection performance when only limited labeled samples are available. For instance, Intxausti et al.\ demonstrated that pre-training on large unlabeled manufacturing X-ray datasets, followed by fine-tuning on a small labeled subset, improves classification accuracy and generalization over purely supervised baselines \cite{lol}.

\subsubsection{Mamba-Based Models for NDT Defect Detection}

The introduction of Mamba, a selective state-space model for efficient modelling of long-range dependencies, has opened a new line of research for NDT image analysis. Mamba and Vision Mamba variants have been incorporated into detection and segmentation architectures for defect imaging, offering linear-time complexity and strong global modelling capabilities.

In object detection, several works have integrated Mamba blocks into YOLO-style detectors. YOLOv5-MDS incorporates Mamba units in the backbone and neck for the detection of printed-circuit-board (PCB) X-ray defects, reporting up to 34\% higher accuracy and 29 to 57\% faster inference than the baseline YOLOv5 model \cite{Guo2025YOLOv5MDS}. A lightweight X-ray-YOLO-Mamba model for prohibited item detection in security X-ray images achieves mAP$_{50\text{–}95}$ values up to 74.6\% at 95\,FPS, demonstrating that Mamba-based detectors can satisfy real-time industrial requirements while improving detection quality \cite{Zhao2025XrayYOLOMamba}. 

For segmentation and surface inspection, Vision Mamba-based architectures such as HMNet, PD-Mamba, and VM-UNet++ have been proposed. HMNet, a high-resolution Mamba network for surface defect segmentation, achieves state-of-the-art accuracy on high-resolution defect datasets while maintaining computational efficiency \cite{Xiao2025HMNet}. PD-Mamba combines Vision Mamba blocks with a robotic inspection platform to detect defects in concrete bridge piers, achieving F1-scores of around 87\% with linear complexity, which is favorable for large-scale infrastructure monitoring \cite{Du2025PDMamba}. VM-U-Net++ and related Mamba-based U-Net variants for crack segmentation report 3–6\% \ improvement in mean Intersection-over-Union (mIoU) over strong CNN and Transformer baselines while using significantly fewer parameters and FLOPs \cite{Tang2025VMUNetPP}.

Hybrid CNN–Mamba U-Nets further demonstrate that combining convolutional branches for local feature extraction with Mamba blocks for global context can yield robust multimodal defect detection. A dual-branch hybrid CNN–Mamba U-Net for coating defect detection, for example, fuses RGB and depth information and outperforms unimodal and non-Mamba baselines on both accuracy and robustness to complex backgrounds \cite{Tang2025DCMUNet}. Taken together, these studies show that Mamba-based sequence modelling is relevant to NDT imaging, especially in high-resolution and real-time scenarios, but the majority of existing applications focus on surface inspection and object detection or segmentation rather than slice-level XCT classification.

\subsection{Motivation and Research Gap}

In the context of XCT inspection for Nomex honeycomb structures, several research gaps can be identified. First, although deep learning has been applied to NDT in related settings, no previous work has presented a fully automated end-to-end pipeline for XCT-based inspection of HCSS components. Industrial inspection of such parts still relies on manual slice-by-slice analysis by expert operators, and XCT images are typically interpreted in isolation from the surrounding production information. As a result, there is no established procedure for extracting raw XCT data from production orders, performing image-level defect classification for Nomex honeycomb structures, and generating standardized outputs that can be directly consumed by digital quality assurance systems.

Second, existing approaches do not provide a complete pipeline to exploit the large volumes of unlabeled historical XCT data routinely generated in manufacturing facilities. Available work on self-supervised or unsupervised learning in NDT often treats datasets as static and does not integrate pretraining, fine-tuning, and deployment into a unified workflow. In particular, the challenges of continual learning and gradual distribution shifts across production orders, such as evolving materials, process parameters, and defect statistics, are rarely addressed in a principled manner, despite their practical importance in long-term industrial deployment.

Third, current Mamba-based NDT models focus primarily on object detection and semantic segmentation of surface or radiographic images, rather than on image-level defect classification of 2D XCT slices. Applications of Mamba or Vision Mamba architectures to XCT honeycomb structures, as well as systematic comparisons against strong CNN baselines in this domain, are essentially absent. Consequently, the potential benefits of Mamba-style state-space modelling and nested learning for slice-level defect classification in industrial XCT remain underexplored.

\subsection{Novelty and Contributions}

This work introduces \emph{NL-MambaXCT}, a nested-learning Mamba-based framework for XCT defect classification in Nomex honeycomb structures. The approach is evaluated on a curated industrial dataset of approximately 2{,}000 expertly labeled XCT slices and a larger pool of approximately 20{,}000 unlabeled slices collected from routine production, reflecting realistic noise, variability, and acquisition artifacts. The main contributions are as follows:

\begin{enumerate}
    \item \textbf{The first fully automated deep-learning pipeline for XCT inspection of Nomex honeycomb structures.}
    This is the first end-to-end framework that processes production-order XCT data and outputs image-level defect classes for direct integration into digital quality-assurance workflows.

    \item \textbf{First integration of nested learning for two-timescale optimization in image-based NDT.} 
    NL-MambaXCT constitutes the first application of the Nested Learning paradigm \cite{behrouz2025nestedlearning} to image-level tasks in general and to non-destructive testing in particular, combining slow EMA traces and fast weights/optimizer states in a unified training scheme. The method is further evaluated with sequential production batches, catastrophic-forgetting metrics, backward transfer, and pre/post adaptation measurements to quantify stability under incremental deployment

    \item \textbf{Mamba-based XCT classifier with self-supervised masked pre-training.}
    A hybrid CNN–Mamba backbone is pre-trained using masked image modelling on unlabeled industrial XCT slices and then fine-tuned on a curated labeled subset, enabling label-efficient defect classification from historical inspection data.

\end{enumerate}

This paper is organized as follows: Section (\S\ref{sec:background}) presents the background on XCT defect classification, Mamba encoders, and nested learning dynamics. Section (\S\ref{sec:method}) details the proposed methodology and the NL-MambaXCT architecture, including pretraining and optimization strategies. Section (\S\ref{sec:experiments}) reports the experimental setup, results, and ablation studies. Finally, Section (\S\ref{sec:conclusion}) concludes the paper and outlines directions for future work.
\section{Background}
\label{sec:background}

\subsection{XCT Defect Classification and Masked Pretraining}
Let $x \in \mathbb{R}^{H \times W}$ denote a single-channel tomographic slice (CT/XCT) and
$y \in \{1,\dots,C\}$ its image-level class label, where $C$ is the number of defect categories.
A labeled dataset is written as
\begin{equation}
\mathcal{D}_{\mathrm{lab}}=\{(x_i,y_i)\}_{i=1}^{N_{\mathrm{lab}}},
\end{equation}
and an unlabeled archive as
\begin{equation}
\mathcal{D}_{\mathrm{unlab}}=\{x^{u}_j\}_{j=1}^{N_u}.
\end{equation}
Given a classifier $f_\theta(\cdot)$ that produces class probabilities $p_\theta(y\mid x)$, supervised
training typically minimizes cross-entropy,
\begin{equation}
\mathcal{L}_{\mathrm{sup}}(\theta)= -\frac{1}{N_{\mathrm{lab}}}\sum_{i=1}^{N_{\mathrm{lab}}}\log p_\theta(y_i\mid x_i).
\end{equation}

When $N_{\mathrm{lab}}$ is limited but $\mathcal{D}_{\mathrm{unlab}}$ is large, masked image modelling (MIM)
is a standard self-supervised strategy that pretrains an encoder by reconstructing masked content from
visible context. Concretely, with a binary mask $M\in\{0,1\}^{H\times W}$ and a mask token/value $t$,
a corrupted view is formed as $\tilde{x}=M\odot x + (1-M)\odot t$, and a reconstruction model
$g_{\theta,\phi}(\cdot)$ predicts $\hat{x}=g_{\theta,\phi}(\tilde{x})$. A common objective emphasizes the
masked region,
\begin{equation}
\mathcal{L}_{\mathrm{MIM}}(\theta,\phi)=
\frac{\| (1-M)\odot(\hat{x}-x)\|_1}{\|(1-M)\|_1+\varepsilon}.
\end{equation}
Importantly, MambaMIM proposes a self-supervised generative pretraining scheme tailored to
Mamba-based vision/medical backbones and highlights that a specific Mamba design is needed for masked
pretraining in state-space models \cite{tang2024mambamim}. The scope of MambaMIM is medical image
analysis (including 3D medical data), rather than industrial NDT \cite{tang2024mambamim}.

\subsection{Mamba Encoders for Image Data}
State-space sequence models of the Mamba family are motivated by efficient long-range dependency
modelling with linear complexity in sequence length. In vision settings, a feature map
$Z\in\mathbb{R}^{B\times D\times H\times W}$ is reshaped into a token sequence
$T\in\mathbb{R}^{B\times N\times D}$, where $B$ is the batch size, $D$ is the feature/channel
dimension, and $N=HW$ is the number of spatial tokens. This representation allows sequence mixing
over spatial tokens while preserving the two-dimensional origin of the XCT slice.
MambaMIM discusses how Mamba and its vision variants have been adopted beyond language, including in
computer vision and medical imaging, and positions its contribution as enabling masked generative
pretraining for Mamba-based backbones \cite{tang2024mambamim}. This prior work is therefore used as a
methodological foundation for Mamba-style encoders and MIM objectives, without attributing any NDT
claims to it.

\subsection{Nested Learning and Two-Timescale Dynamics}
Nested Learning (NL) presented in \cite{behrouz2025nestedlearning} reformulates learning systems as collections of interacting optimization problems
operating at different update frequencies (fast and slow ``levels''), each with its own context flow. In this view, optimizers can themselves be interpreted as learning
modules; for example, the NL paper revisits momentum SGD via the standard recursion
\begin{align}
W_{i+1} &= W_i + m_{i+1}, \\
m_{i+1} &= \alpha_{i+1} m_i - \eta_i \nabla \mathcal{L}(W_i; x_i),
\end{align}
and argues that the momentum state behaves as a form of memory over gradients \cite{behrouz2025nestedlearning}.
More generally, NL introduces an explicit notion of update frequency and orders components into levels
based on how often their states change. This conceptual framework
motivates two-timescale mechanisms in settings where data distributions drift over time and continual
updates are expected, while the specific instantiation used in the proposed method is described in
Section~\ref{sec:method}.
\begin{figure*}[!h]
    \centering
    \includegraphics[width=\linewidth]{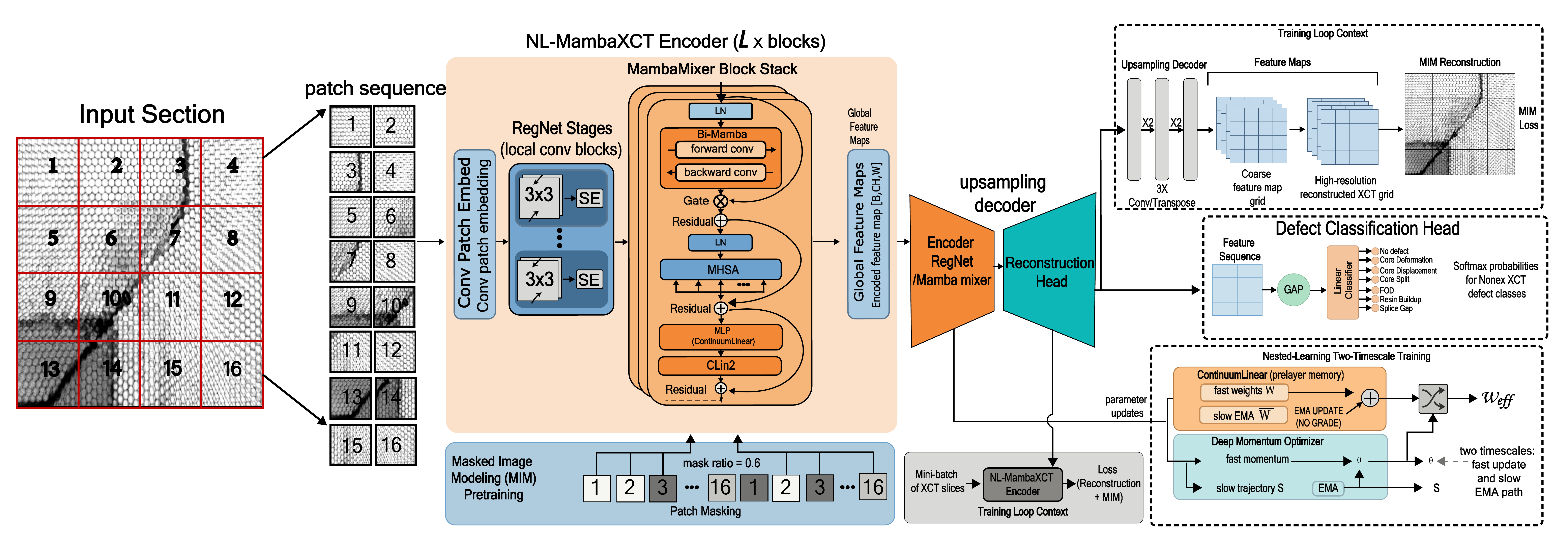}
    \caption{NL-MambaXCT training pipeline for Nomex honeycomb X-ray CT defect classification. The input XCT slice is patchified and processed by a detailed NL-Mamba encoder, jointly optimized via masked image modeling (MIM) reconstruction (top branch) and a defect classification head (bottom branch), while a nested-learning two-timescale scheme (ContinuumLinear fast/slow weights + deep momentum optimizer) updates the model parameters.}
    \label{fig:graphical-abstract}
\end{figure*}
\section{Methodology}
\label{sec:method}

This section presents the NL-MambaXCT methodology. The methodological emphasis is placed on the proposed \emph{Nested Learning instantiation} that introduces explicit fast/slow parameter dynamics at both the layer level and the optimizer level, and on how these dynamics are integrated into an industrial XCT training and deployment workflow. Figure~\ref{fig:graphical-abstract} provides a graphical overview of the pipeline.

\subsection{Data Acquisition and Preprocessing}

\begin{figure}[h!]
\centering
\includegraphics[width=\linewidth]{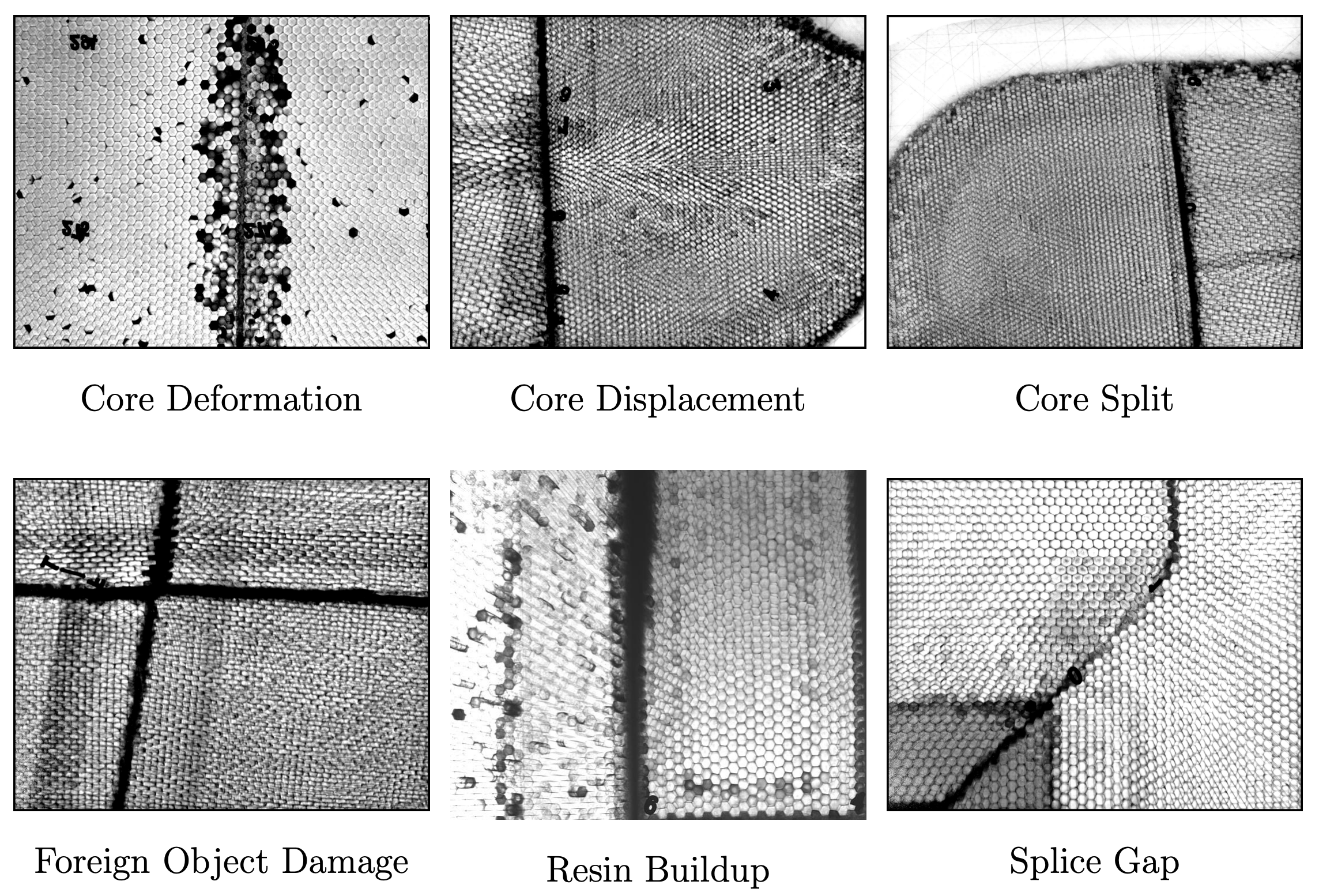}
\caption{Types of defects detected by X-ray computed tomography for the Nomex honeycomb structures }
\label{fig:defects}
\end{figure}

XCT data are collected from Strata Manufacturing PJSC. This leading aerospace composites manufacturer supplies structural components to major aircraft manufacturers. The dataset consists of (i) an unlabeled historical archive acquired during routine production and (ii) a smaller curated subset with technician-verified defect labels. Each sample corresponds to a single 2D grayscale XCT slice of a Nomex honeycomb component. The labeled subset is curated to remove low-quality scans and ambiguous cases. The types of defects are depicted in Figure~\ref{fig:defects}.

All slices are converted to a single-channel representation and spatially standardized using resizing and center-cropping to a fixed resolution. Intensity normalization is applied to reduce scanner- and batch-dependent contrast variation. During training, lightweight geometric augmentations are applied, including random horizontal flipping. These augmentations improve robustness to modest slice-position and in-plane orientation differences, while the production-order split evaluates the model on later scans that were not used during training. Larger out-of-plane rotations or substantially different scanner setups remain deployment conditions that should be covered by additional calibration data before use in a new inspection configuration.

\subsection{Problem Formulation and Training Stages}

The task is posed as image-level multi-class defect classification on XCT slices. The training is performed in two stages. First, self-supervised MIM pretraining is performed on the unlabeled archive to learn representations tailored to the industrial XCT domain. Second, supervised fine-tuning is performed on the curated labeled subset. The distinctive component of NL-MambaXCT is that fine-tuning is not treated as a single-timescale optimization process; instead, nested-learning dynamics are instantiated through (i) layer-level fast/slow linear projections and (ii) an optimizer with an explicit slow parameter trajectory. These mechanisms are detailed in Sections~\ref{sec:continuum} and~\ref{sec:deep_momentum}.

\subsection{NL-MambaXCT Encoder: Hybrid CNN--Mamba Backbone}

A hybrid CNN--Mamba encoder is employed to balance local defect sensitivity with global context modelling. This design replaces a purely CNN backbone (e.g., ResNet-50 end-to-end classification) with a structure in which early stages are convolutional while deeper stages incorporate sequence-mixing operators. The encoder is organized as a four-stage hierarchy followed by global pooling and a linear classifier head.

\subsubsection{Stages 1--2: RegNet Convolutional Blocks}

The first two stages employ lightweight RegNet bottleneck blocks to learn robust local features. These stages are retained explicitly to preserve sensitivity to fine-scale radiographic structures, which are essential in honeycomb XCT where many defects manifest as subtle changes in local density and texture.

\subsubsection{Stages 3--4: MambaMixer Blocks}

The last two stages apply sequence mixing over the patch grid. In NL-MambaXCT, fast and slow dynamics of nested learning are embedded within the projection layers used by the mixer and the classification head. Concretely, the linear projections used for channel mixing and gating are replaced by the proposed fast/slow linear projections described in Section~\ref{sec:continuum}. This design enables the deeper sequence-mixing stages to adapt rapidly through fast weights while being regularized by a slowly evolving parameter trace.

\subsection{Masked Image Modeling Pretraining}

Self-supervised pretraining follows a masked image modelling objective,  a fixed fraction of patches is masked, a reconstruction module predicts the missing content, and the encoder is optimized to minimize a masked reconstruction loss. The formulation of the masking operator, mask token injection, decoder structure, and reconstruction loss are discussed in the Background (Section~\ref{sec:background}) and are therefore not repeated here.

The methodological significance in the present context is that MIM is applied to a large, unlabeled XCT archive originating from routine production. This archive captures scanner variation, component variation, and operational drift that are typically absent from public benchmarks. Pretraining on this distribution is used to reduce reliance on scarce labeled samples and to align the encoder with domain-specific radiographic statistics before supervised fine-tuning.

\subsection{Continuum Linear Projection: Layer-Level Nested Two-Timescale Dynamics}
\label{sec:continuum}

Nested Learning (NL) posits that continual learning and stabilized adaptation can be induced by organizing learning dynamics across multiple timescales. In NL-MambaXCT, this principle is instantiated at the level of linear projections by maintaining (i) \emph{fast} trainable weights updated by backpropagation and (ii) a \emph{slow} parameter trace updated by a separate rule. Gradients are not propagated through the slow trace, which enforces a strict separation between fast adaptation and slow consolidation.

Consider a linear projection with input $\mathbf{h}\in\mathbb{R}^{d_{\mathrm{in}}}$ and output $\mathbf{u}\in\mathbb{R}^{d_{\mathrm{out}}}$. Two parameter states are maintained at the optimization step $t$:
(i) fast parameters $(\mathbf{W}_t,\mathbf{b}_t)$ and
(ii) slow parameters $(\bar{\mathbf{W}}_t,\bar{\mathbf{b}}_t)$.
The slow parameters are updated using an exponential moving average (EMA) of the fast parameters:
\begin{align}
\bar{\mathbf{W}}_{t+1} &= (1-\alpha)\,\bar{\mathbf{W}}_{t} \;+\; \alpha\,\operatorname{sg}(\mathbf{W}_{t}),
\label{eq:slowW_update}\\
\bar{\mathbf{b}}_{t+1} &= (1-\alpha)\,\bar{\mathbf{b}}_{t} \;+\; \alpha\,\operatorname{sg}(\mathbf{b}_{t}),
\label{eq:slowb_update}
\end{align}
where $\alpha\in(0,1)$ controls the slow time scale and $\operatorname{sg}(\cdot)$ denotes a stop-gradient operator (i.e., identity in the forward pass and zero derivative in the backward pass). This construction ensures that the slow state evolves without receiving gradients.

During the forward pass, an \emph{effective} parameterization is used that blends fast parameters with the slow trace:
\begin{align}
\tilde{\mathbf{W}}_{t} &= \mathbf{W}_{t} \;+\; \lambda\,\operatorname{sg}\!\bigl(\bar{\mathbf{W}}_{t}-\mathbf{W}_{t}\bigr),
\label{eq:W_eff}\\
\tilde{\mathbf{b}}_{t} &= \mathbf{b}_{t} \;+\; \lambda\,\operatorname{sg}\!\bigl(\bar{\mathbf{b}}_{t}-\mathbf{b}_{t}\bigr),
\label{eq:b_eff}
\end{align}
with $\lambda\in[0,1]$ controlling the influence of the slow trace on the forward computation. The projection output is then computed as
\begin{equation}
\mathbf{u} \;=\; \tilde{\mathbf{W}}_{t}\mathbf{h} \;+\; \tilde{\mathbf{b}}_{t}.
\label{eq:clp_forward}
\end{equation}

Equations~\eqref{eq:slowW_update}--\eqref{eq:clp_forward} instantiate a two-time scale learning rule in which the fast parameters $\mathbf{W}_t$ receive gradients and can rapidly adapt to the labeled supervision signal, while the slow trace $(\bar{\mathbf{W}}_t,\bar{\mathbf{b}}_t)$ acts as a stabilizing memory that integrates updates over a longer horizon. Since the slow trace is injected only through a stop-gradient path, the learning problem is not converted into a higher-order optimization objective; instead, it remains first-order while still expressing a nested fast/slow structure consistent with NL principles.

In NL-MambaXCT, these projections are used in (i) the deep sequence-mixing stages and (ii) the classification head. This placement is intentional: deeper features are most sensitive to distribution drift across production orders, and the slow trace provides regularization against abrupt representation shifts when labels are limited and class imbalance is severe.

\subsection{Deep Momentum SGD: Optimizer-Level Nested Dynamics}
\label{sec:deep_momentum}

A second nested timescale is introduced at the optimizer level by maintaining a slow parameter trajectory in addition to standard momentum dynamics. This mechanism can be interpreted as an optimizer-level analogue of the layer-level slow trace: the fast iterate follows the gradient signal, while the slow iterate integrates the fast trajectory over time and is mixed back into the update to damp oscillations.

Let $\boldsymbol{\theta}_t$ denote the collection of trainable parameters in the optimization step $t$, and let $\mathbf{g}_t=\nabla_{\boldsymbol{\theta}}\mathcal{L}_t$ denote the gradient of training loss in that step. The standard momentum is maintained through a velocity vector $\mathbf{v}_t$:
\begin{equation}
\mathbf{v}_{t+1} \;=\; \mu\,\mathbf{v}_t \;+\; \mathbf{g}_t,
\label{eq:momentum_update}
\end{equation}
with momentum coefficient $\mu\in[0,1)$. A fast parameter proposal is produced as
\begin{equation}
\boldsymbol{\theta}^{\mathrm{fast}}_{t+1} \;=\; \boldsymbol{\theta}_t \;-\; \eta\,\mathbf{v}_{t+1},
\label{eq:theta_fast}
\end{equation}
where $\eta>0$ is the learning rate.

A slow trajectory $\mathbf{s}_t$ is then updated as an EMA of the fast proposal:
\begin{equation}
\mathbf{s}_{t+1} \;=\; \rho\,\mathbf{s}_t \;+\; (1-\rho)\,\operatorname{sg}\!\bigl(\boldsymbol{\theta}^{\mathrm{fast}}_{t+1}\bigr),
\label{eq:slow_traj}
\end{equation}
with $\rho\in(0,1)$ controlling the slow time scale. Finally, the actual parameter update is defined as a convex combination of the fast proposal and the slow trajectory:
\begin{equation}
\boldsymbol{\theta}_{t+1} \;=\; (1-\gamma)\,\boldsymbol{\theta}^{\mathrm{fast}}_{t+1} \;+\; \gamma\,\mathbf{s}_{t+1},
\qquad \gamma\in[0,1].
\label{eq:theta_mix}
\end{equation}

Equations~\eqref{eq:momentum_update}--\eqref{eq:theta_mix} introduce an explicit optimizer-level nesting: the fast iterate reacts immediately to minibatch gradients, while the slow trajectory aggregates information over many steps and is injected back into the parameter stream. In practice, this reduces sensitivity to noisy gradients and stabilizes training dynamics in small, imbalanced industrial datasets.

When combined with the layer-level mechanism in Section~\ref{sec:continuum}, a nested structure of two-levels is obtained: There are slow traces within selected linear projections and an additional slow trajectory exists at the optimizer level. This design is aligned with the central premise of NL that different components can be updated at different effective frequencies, enabling continuous stable adaptation without the need to backpropagate through slow states.

\subsection{Supervised Fine-Tuning Objective}

After MIM pretraining, the reconstruction head is discarded and the encoder is fine-tuned for defect classification on the curated labeled dataset. The supervised objective is the standard multi-class cross-entropy loss, optionally with class reweighting to mitigate imbalance. Training is performed with early stopping based on the validation macro F1-score. The nested-learning mechanisms described above remain active during fine-tuning: slow traces are updated by EMA rules and are used in forward computation via stop-gradient blending, while the optimizer maintains both fast momentum and a slow parameter trajectory.
\section{Experiments and Results}
\label{sec:experiments}

This section evaluates NL-MambaXCT on industrial Nomex honeycomb XCT data and analyzes the contribution of each component of the proposed model. The experiments are designed to address four questions: \begin{itemize}
    \item How NL-MambaXCT compares to strong CNN and Mamba baselines on the curated Nomex dataset;
    \item To what extent self-supervised masked image modelling (MIM) pretraining on unlabeled slices improves performance; and
    \item How much the nested-learning components contribute relative to a non-nested Mamba backbone; and
    \item Whether sequential production-batch updates lead to catastrophic forgetting or stable adaptation under distribution drift.
\end{itemize}
 All training is performed on an NVIDIA DGX system equipped with 8 A100 40\,GB GPUs, with a single GPU used per run unless otherwise specified. The implementation follows the training procedures described in the MIM pretraining and fine-tuning scripts, as well as the Nested-XCT-Mamba backbone definition. 

\subsection{Dataset and Experimental Setup}
\label{subsec:setup}

\subsubsection{Labeled dataset}

Supervised experiments use a curated labeled dataset of 2D XCT slices from Nomex honeycomb sandwich structures. Starting from an internal archive, noisy scans, duplicated views, and ambiguous cases are removed in collaboration with certified X-ray inspectors, resulting in a final dataset of 2{,}000 slices. Each slice is assigned to exactly one of seven mutually exclusive classes: \emph{No Defect}, \emph{Core Deformation}, \emph{Core Displacement}, \emph{Core Split}, \emph{Foreign Object Damage (FOD)}, \emph{Resin Buildup}, and \emph{Splice Gap}. The class distribution is summarised in Table~\ref{tab:labeled_distribution}. 

\begin{table*}[t]
\centering
\begin{minipage}[t]{0.47\linewidth}
\centering
\caption{Final distribution of labeled XCT slices per defect class.}
\label{tab:labeled_distribution}
\footnotesize
\setlength{\tabcolsep}{8pt}
\renewcommand{\arraystretch}{1.15}
\begin{tabular}{@{}lc@{}}
\toprule
\rowcolor{gray!10}
\textbf{Defect class} & \textbf{\# Images} \\
\midrule
\rowcolor{gray!3}
No Defect                     &  780 \\
Core Deformation              &  211 \\
\rowcolor{gray!3}
Core Displacement             &  213 \\
Core Split                    &  187 \\
\rowcolor{gray!3}
Foreign Object Damage (FOD)   &  210 \\
Resin Buildup                 &  195 \\
\rowcolor{gray!3}
Splice Gap                    &  204 \\
\midrule
\textbf{Total}                & \textbf{2{,}000} \\
\bottomrule
\end{tabular}
\end{minipage}\hfill
\begin{minipage}[t]{0.49\linewidth}
\centering
\caption{Train/validation/test splits based on production orders.}
\label{tab:splits}
\footnotesize
\setlength{\tabcolsep}{8pt}
\renewcommand{\arraystretch}{1.15}
\begin{tabular}{@{}lccc@{}}
\toprule
\rowcolor{gray!10}
\textbf{Split} & \textbf{\# Orders} & \textbf{\# Slices} & \textbf{Fraction} \\
\midrule
\rowcolor{gray!3}
Train      & 32 & 1{,}300 & 65\% \\
Validation &  8 &   200  & 10\% \\
\rowcolor{gray!3}
Test       & 10 &   500  & 25\% \\
\bottomrule
\end{tabular}
\end{minipage}
\end{table*}

\subsubsection{Unlabeled pretraining pool}

For self‑supervised pretraining, a separate archive of  19{,}961 unlabeled XCT slices collected from routine production is used. These slices originate from multiple production orders, material batches, and scanning sessions, thereby capturing realistic variability in acquisition conditions and defect prevalence. No manual labels are required at this stage; the unlabeled pool is used exclusively for MIM pretraining of the encoder. 

\subsubsection{Production‑order splits}

To mimic a realistic deployment scenario, train/validation/test splits are defined at the level of \emph{production orders} rather than individual slices. Each production order corresponds to a single XCT scan of a Nomex honeycomb component, from which multiple 2D slices are extracted around the region of interest. All slices originating from the same production order are assigned to the same split, based on the order’s scanning date: the earliest orders are used for training, an intermediate set of orders is used for validation, and the most recent orders are reserved for testing. This chronological splitting strategy prevents leakage of slices from the same component across splits and approximates the situation of deploying the model on future production.

The resulting split statistics are shown in Table~\ref{tab:splits}; the number of slices is larger than the number of orders because several slices are extracted per component. The class distribution and the production-order splits are summarised together in Tables~\ref{tab:labeled_distribution} and~\ref{tab:splits}.

\subsubsection{Preprocessing, augmentations, and training}

All slices are converted to single–channel images, resized and centre–cropped to $256{\times}256$ pixels. During MIM pretraining, the images are normalized to zero mean and unit variance, and standard geometric augmentations (random horizontal flips and random resized crops) are applied to improve robustness to minor viewpoint variations. During supervised fine–tuning and evaluation, a lighter augmentation pipeline is used: random resized cropping and horizontal flipping for the training set, and deterministic resizing followed by centre cropping for the validation and test sets.

For self–supervised MIM pretraining, NL–MambaXCT is optimized in the unlabeled pool for $T_{\text{pre}}$ epochs using the AdamW optimizer with a batch size of 32, a learning rate $10^{-5}$, weight decay $0.005$, and a mask ratio of $0.6$, the fraction of patches masked per image. The objective is the masked-region reconstruction loss defined in Section~\ref{sec:background}, computed between the original and reconstructed pixels on the masked patches.

For supervised fine–tuning, the encoder parameters are initialized from the MIM-pre-trained weights, and a linear classification head is attached on top of the global pooled representation. The model is then trained for $T_{\text{ft}}$ epochs in the labeled Nomex dataset using Deep Momentum SGD with a batch size of 32, a learning rate $10^{-5}$ and a weight decay $10^{-4}$. All competing models (CNN baselines, non–nested Mamba backbone and NL-MambaXCT variants) follow the same training/validation protocol, differing only in architecture and optimizer choice. Early stopping based on the F1-score validation macro is used to select the final checkpoint for each model. In the held–out test set, overall accuracy, macro–averaged precision, recall, and F1–score are reported, together with per–class precision, recall, and F1–scores.

\subsubsection{Sequential production-batch continual-learning protocol}
\label{sec:seq_cl_protocol}

To evaluate catastrophic forgetting and adaptation to production drift, an additional sequential-batch protocol is conducted on a dedicated operational continual-learning dataset. This dataset is assembled separately from the 2,000-slice classification dataset in Section~\ref{subsec:setup}, so it is not a subset of the labeled classification dataset but rather has the same demographics as the original dataset.

The model is initialized from the strongest fine-tuned NL-MambaXCT checkpoint. At step $t$, the model receives production batch $B_t$ and is updated for 15 epochs with learning rate of $10^{-4}$, a 0.1 backbone learning-rate multiplier, a 20.0 head learning-rate multiplier, weighted cross-entropy, and full replay of previously observed batches. This setting emulates a conservative industrial update policy in which new production data are incorporated without discarding earlier labeled evidence. As a sanity-check baseline, the same stream is also evaluated with a naive plain-Mamba sequential update that disables the ContinuumLinear fast/slow nested traces and does not replay earlier batches.

Let $F_{t,b}$ denote the macro F1-score on production batch $B_b$ after learning through step $t$. The continual-learning evaluation reports the following metrics; these abbreviations are used in the corresponding results tables:
\begin{itemize}
    \item \textbf{Test Acc.} and \textbf{Test Macro-F1}: accuracy and macro F1 on the test set after the final production-batch update.
    \item \textbf{Avg-Batch Macro-F1}: average macro F1 across all four production batches after the final update, measuring retained batch-level performance.
    \item \textbf{MF} and \textbf{MaxF}: mean and maximum catastrophic forgetting, defined as the decrease from the best previously observed performance on an old batch to its final performance,
\begin{equation}
\mathrm{Forgetting}_b = \max_{t \leq T} F_{t,b} - F_{T,b}, \qquad b<T .
\end{equation}
    Lower values indicate better retention.
    \item \textbf{BWT}: backward transfer, measured as the final change on old batches, $F_{T,b}-F_{b,b}$ for $b<T$. Positive values indicate that later updates improve earlier batches on average, while negative values indicate forgetting.
    \item \textbf{IAG}: incoming-batch adaptation gain, measured as the improvement from pre-update to post-update performance on each incoming production batch, $F_{t,t}-F_{t-1,t}$ for $t>0$. This directly evaluates adaptation to distribution drift.
\end{itemize}
These metrics complement the global held-out test accuracy and macro F1-score by explicitly tracking stability on old production batches and plasticity on newly introduced batches.

\subsection{Comparison with Baseline Models}
\label{subsec:main_results}

The first set of experiments compares NL-MambaXCT with representative supervised baselines on the Nomex XCT dataset. The baselines include a ResNet-50 classifier initialized from ImageNet weights, a RegNetY-style CNN with capacity similar to the convolutional stem of NL-MambaXCT, an EfficientNet variant with comparable parameter count, ViT-B/16 and Swin-T attention baselines initialized from ImageNet weights, and a plain MambaXCT backbone in which all ContinuumLinear layers and Deep Momentum components are disabled. The plain Mamba model isolates the effect of the Mamba-style sequence mixer without nested learning or MIM pretraining, while ViT-B/16 and Swin-T provide explicit attention-based comparisons requested for this revision. A dedicated computational benchmark in Table~\ref{tab:efficiency} compares against ResNet-50, ViT-B/16, and Swin-T, quantifying the efficiency of the proposed state-space design relative to convolutional, global-attention, and windowed-attention model families. All evaluated classification models use the same training/validation/test split, batch size, image resolution, and augmentation policy where applicable, and are evaluated on the same production-order test set.

Table~\ref{tab:main_results} summarises the test performance. The NL‑MambaXCT model pretrained with MIM on the 19{,}961 unlabeled slices and subsequently fine‑tuned on the labeled dataset achieves an accuracy of 96.8\% and a macro F1‑score of 96.7\%, outperforming all competing baselines by a margin between 4.1 and 10.2 percentage points in accuracy (the smallest margin being over the strongest baseline, ResNet-50). The best CNN baseline (ResNet‑50) achieves 92.7\% accuracy and 92.6\% macro F1, while the plain MambaXCT backbone without nested learning reaches 92.2\% accuracy. The ViT-B/16 and Swin-T attention baselines reach 86.6\% and 86.8\% accuracy, respectively, on the same held-out test split, confirming that the improvement is not simply a consequence of using a generic attention backbone. Training NL-MambaXCT from scratch without MIM pretraining yields 93.8\% accuracy, indicating that both the architectural changes and the self-supervised pretraining contribute to the observed improvements.

\begin{table*}[h!]
\centering
\caption{Final test-set performance on the Nomex XCT dataset using the production-order split.}
\label{tab:main_results}
\footnotesize
\setlength{\tabcolsep}{5.5pt}
\renewcommand{\arraystretch}{1.08}
\begin{tabular}{@{}lcccc@{}}
\toprule
\rowcolor{gray!10}
\textbf{Model} & \textbf{Acc.} & \textbf{Macro Prec.} & \textbf{Macro Rec.} & \textbf{Macro F1} \\
\midrule
\rowcolor{gray!3}
ResNet-50                 & 92.7\% & 92.9\% & 92.4\% & 92.6\% \\
RegNetY                   & 91.5\% & 91.8\% & 91.0\% & 91.4\% \\
\rowcolor{gray!3}
EfficientNet              & 89.9\% & 90.3\% & 89.2\% & 89.7\% \\
ViT-B/16 (ImageNet-pretrained) & 86.6\% & 85.1\% & 84.9\% & 84.7\% \\
\rowcolor{gray!3}
Swin-T (ImageNet-pretrained) & 86.8\% & 85.5\% & 84.3\% & 84.8\% \\
Plain MambaXCT & 92.2\% & 92.5\% & 91.9\% & 92.1\% \\
\rowcolor{gray!3}
NL-MambaXCT (no MIM)     & 93.8\% & 94.1\% & 93.4\% & 93.7\% \\
NL-MambaXCT (MIM-pretrained) & \best{96.8\%} & \best{96.7\%} & \best{96.7\%} & \best{96.7\%} \\
\bottomrule
\end{tabular}
\end{table*}

A per-class breakdown for the MIM-pretrained NL-MambaXCT model is provided in Table~\ref{tab:per_class_results}. Performance is consistently high across all seven classes, including the visually subtle \emph{Core Deformation} and \emph{Core Displacement} categories, which historically have been challenging due to mild contrast against the background honeycomb cells. The model maintains F1‑scores above 0.95 for all classes, with slightly higher values for more visually distinct defects such as FOD and No Defect.

\begin{table}[h!]
\centering
\caption{Per-class metrics for NL-MambaXCT (MIM-pretrained) on the test set.}
\label{tab:per_class_results}
\footnotesize
\setlength{\tabcolsep}{4.5pt}
\renewcommand{\arraystretch}{1.08}
\begin{tabular}{@{}lccc@{}}
\toprule
\rowcolor{gray!10}
\textbf{Class} & \textbf{Precision} & \textbf{Recall} & \textbf{F1-Score} \\
\midrule
\rowcolor{gray!3}
No Defect         & \best{0.971} & \best{0.971} & \best{0.971} \\
Core Deformation  & 0.980 & 0.961 & 0.970 \\
\rowcolor{gray!3}
Core Displacement & 0.980 & 0.962 & 0.971 \\
Core Split        & 0.956 & 0.956 & 0.956 \\
\rowcolor{gray!3}
FOD               & 0.962 & 0.980 & 0.971 \\
Resin Buildup     & 0.957 & 0.957 & 0.957 \\
\rowcolor{gray!3}
Splice Gap        & 0.961 & 0.980 & 0.970 \\
Macro average & 0.967 & 0.967 & 0.967 \\
\bottomrule
\end{tabular}
\end{table}

\begin{figure}[h!]
    \centering
    \includegraphics[width=\linewidth]{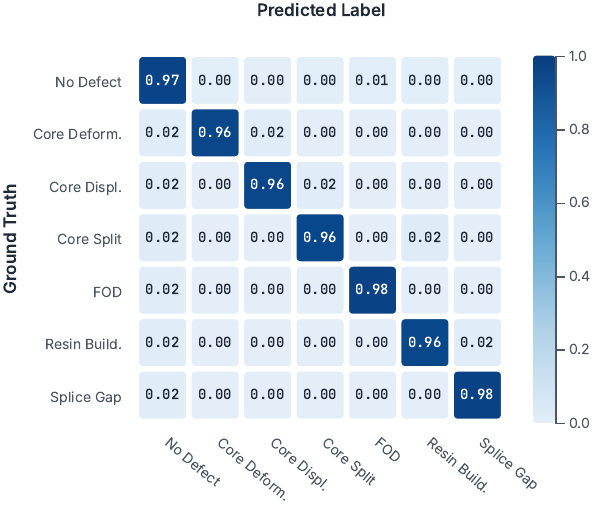}
    \caption{Normalised confusion matrix for NL-MambaXCT (MIM-pretrained) on the held-out test set.}
    \label{fig:nl_mambaxct_confusion}
\end{figure}

Figure~\ref{fig:nl_mambaxct_confusion} depicts the normalized confusion matrix for NL-MambaXCT, showing that misclassifications are rare and scattered rather than concentrated in any single class pair. The residual errors mostly involve the \emph{No Defect} class and the visually subtle \emph{Core Deformation}, \emph{Core Displacement}, and \emph{Core Split} categories, together with a small \emph{Resin Buildup}--\emph{Splice Gap} confusion; in each case only one or two slices per class are affected. These confusions are consistent with the slice-level formulation: subtle defects can resemble defect-free honeycomb in an individual cross-section, and for defects that extend obliquely through the volume the discriminative evidence may appear only in adjacent slices. The low off-diagonal values indicate that the model reliably separates the categories in the vast majority of cases, while the remaining errors occur near ambiguous transition slices where a single 2D cross-section does not fully capture the three-dimensional defect morphology.

\subsubsection{Continual Learning under Sequential Production Batches}

Table~\ref{tab:cl_summary} reports the sequential production-batch continual-learning metrics, together with a naive sequential-update sanity check. The final NL-MambaXCT model reaches 97.07\% test accuracy and 96.64\% test macro F1-score on the held-out test set after all four production batches have been incorporated. More importantly for continual learning, the final average macro F1-score across the four production batches is 97.89\%, with a mean forgetting of only 0.08\% and a maximum forgetting of 0.21\%. Backward transfer is slightly positive (+0.11\%), indicating that later updates do not degrade earlier batches on average. The mean post-update adaptation gain on incoming batches is +0.87\%, showing that the model improves on newly introduced production batches while preserving previous knowledge. In contrast, the naive plain-Mamba sequential baseline has lower held-out macro F1 (96.08\%), lower final average production-batch macro F1 (97.33\%), larger mean forgetting (0.60\%), larger maximum forgetting (1.52\%), and negative backward transfer (-0.60\%).

\begin{table}[ht]
\centering
\caption{Sequential production-batch continual-learning metrics on the operational continual-learning dataset. Metric definitions are given in Section~\ref{sec:seq_cl_protocol}.}
\label{tab:cl_summary}
\footnotesize
\setlength{\tabcolsep}{6pt}
\renewcommand{\arraystretch}{1.15}
\begin{tabular}{@{}lcc@{}}
\toprule
\rowcolor{gray!10}
\textbf{Metric} & \textbf{\shortstack{NL-MambaXCT\\(+ replay)}} & \textbf{\shortstack{Naive plain\\Mamba}} \\
\midrule
\rowcolor{gray!3}
Test Acc. & \best{97.07\%} & 96.48\% \\
Test Macro-F1 & \best{96.64\%} & 96.08\% \\
\rowcolor{gray!3}
Avg-Batch Macro-F1 & \best{97.89\%} & 97.33\% \\
MF $\downarrow$ & \best{0.08\%} & 0.60\% \\
\rowcolor{gray!3}
MaxF $\downarrow$ & \best{0.21\%} & 1.52\% \\
BWT & \best{+0.11\%} & -0.60\% \\
\rowcolor{gray!3}
IAG & \best{+0.87\%} & +0.79\% \\
\bottomrule
\end{tabular}
\end{table}

The batch-wise macro F1 matrix in Table~\ref{tab:cl_matrix} shows how NL-MambaXCT changes after each sequential update. Rows correspond to the model after learning a production batch, while columns correspond to evaluation on each production batch. The final row shows that all four batches retain macro F1-scores between 97.43\% and 98.23\%, despite being learned sequentially.  Table~\ref{tab:cl_naive_matrix} summarizes the final retained batch performance and per-batch forgetting for both methods. The largest practical difference occurs on Batch 2 and Batch 4: NL-MambaXCT retains 1.04 and 1.05 percentage points higher macro F1, respectively, and reduces Batch 2 forgetting from 1.52\% to 0.00\%. This directly addresses catastrophic forgetting by showing that previously learned production batches remain more stable under the proposed nested-learning update with replay than under the naive sequential update.

\begin{table*}[t]
\centering
\begin{minipage}[t]{0.46\linewidth}
\centering
\begin{minipage}[t][4.8\baselineskip][t]{\linewidth}
\caption{Batch-wise macro F1-score matrix for NL-MambaXCT with replay under the sequential production-batch protocol. Rows indicate the model state after each update; entries above the diagonal are pre-adaptation evaluations on future incoming batches.}
\label{tab:cl_matrix}
\end{minipage}
\par
\scriptsize
\setlength{\tabcolsep}{4pt}
\renewcommand{\arraystretch}{1.25}
\begin{tabular}{@{}lcccc@{}}
\toprule
\rowcolor{gray!10}
\textbf{Model state} & \textbf{Batch 1} & \textbf{Batch 2} & \textbf{Batch 3} & \textbf{Batch 4} \\
\midrule
\rowcolor{gray!3}
After batch 1 & 96.90\% & 97.23\% & 97.54\% & 96.63\% \\
After batch 2 & \best{97.46\%} & \best{98.09\%} & 97.55\% & 97.10\% \\
\rowcolor{gray!3}
After batch 3 & 97.12\% & \best{98.09\%} & \best{98.44\%} & 96.95\% \\
After batch 4 & 97.43\% & \best{98.09\%} & 98.23\% & \best{97.81\%} \\
\bottomrule
\end{tabular}
\end{minipage}\hfill
\begin{minipage}[t]{0.5\linewidth}
\centering
\begin{minipage}[t][4.8\baselineskip][t]{\linewidth}
\caption{Final retained batch performance and forgetting comparison after all sequential updates. Forgetting is reported only for previously learned batches; Batch 4 has no later update and therefore no forgetting entry.}
\label{tab:cl_naive_matrix}
\end{minipage}
\par
\scriptsize
\setlength{\tabcolsep}{4pt}
\renewcommand{\arraystretch}{1.25}
\begin{tabular}{@{}lccccc@{}}
\toprule
\rowcolor{gray!10}
\textbf{Batch} & \textbf{\shortstack{NL\\final F1}} & \textbf{\shortstack{Naive\\final F1}} & \textbf{$\Delta$ F1} & \textbf{\shortstack{NL\\F$\downarrow$}} & \textbf{\shortstack{Naive\\F$\downarrow$}} \\
\midrule
\rowcolor{gray!3}
Batch 1 & \best{97.43\%} & 97.41\% & +0.02\% & \best{0.03\%} & 0.05\% \\
Batch 2 & \best{98.09\%} & 97.05\% & \best{+1.04\%} & \best{0.00\%} & 1.52\% \\
\rowcolor{gray!3}
Batch 3 & \best{98.23\%} & 98.10\% & +0.13\% & \best{0.21\%} & 0.23\% \\
Batch 4 & \best{97.81\%} & 96.76\% & \best{+1.05\%} & -- & -- \\
\bottomrule
\end{tabular}
\end{minipage}
\end{table*}

These results support the intended role of the nested-learning formulation in deployment: the model can be adapted to new production batches without losing performance on earlier batches. The sequential protocol is deliberately stricter than a single static train/test split because it evaluates the model repeatedly as the training distribution evolves. The near-zero forgetting, positive backward transfer, and measurable adaptation gain indicate that the proposed framework is suitable for incremental industrial use where future production orders may differ slightly from the original labeled archive.

\subsubsection{Computational Efficiency}

Because industrial XCT inspection must support high-throughput production review, computational efficiency is interpreted together with predictive performance rather than as a raw latency contest. The hybrid design of NL-MambaXCT is intended to retain local convolutional sensitivity in early stages while using linear-complexity Mamba sequence mixing in deeper stages, avoiding the quadratic token-token interaction cost of global self-attention. Table~\ref{tab:efficiency} therefore reports accuracy, macro F1, parameter count, profiled floating-point operations (FLOPs), and single-image CPU throughput for the proposed model and representative CNN/Transformer baselines. The plain MambaXCT row is included only as an internal ablation: it is faster because it removes the nested traces, but it also loses accuracy and the stability benefits shown in Table~\ref{tab:nl_ablation}. Profiling was conducted with batch size 1 on a $256\times256$ single-channel XCT input using the same local PyTorch profiler configuration for all models.

\begin{table*}[h!]
\centering
\caption{Accuracy--efficiency tradeoff for the proposed model and representative baselines. The highlighted cells mark the strongest predictive performance, while the plain MambaXCT row is an internal ablation rather than the final method.}
\label{tab:efficiency}
\scriptsize
\setlength{\tabcolsep}{4.5pt}
\renewcommand{\arraystretch}{1.08}
\begin{tabular}{@{}lcccccc@{}}
\toprule
\rowcolor{gray!10}
\textbf{Model} & \textbf{Role} & \textbf{Acc.} & \textbf{Macro F1} & \textbf{Params (M) $\downarrow$} & \textbf{FLOPs (G) $\downarrow$} & \textbf{CPU FPS} \\
\midrule
\rowcolor{gray!3}
ResNet-50 & CNN baseline & 92.7\% & 92.6\% & 23.514 & 10.471 & 4.38 \\
ViT-B/16 & Attention baseline & 86.6\% & 84.7\% & 85.456 & 46.197 & 1.34 \\
\rowcolor{gray!3}
Swin-T & Attention baseline & 86.8\% & 84.8\% & 27.521 & 14.677 & 0.90 \\
Plain MambaXCT & Internal ablation & 92.2\% & 92.1\% & 9.027 & \best{2.557} & \best{13.42} \\
\rowcolor{gray!3}
NL-MambaXCT & Final method & \best{96.8\%} & \best{96.7\%} & \best{9.027} & 2.567 & 11.01 \\
\bottomrule
\end{tabular}
\end{table*}

The profiling results confirm that NL-MambaXCT gives the strongest accuracy--efficiency tradeoff. Compared with ResNet-50, NL-MambaXCT improves accuracy by 4.1 percentage points while using 61.6\% fewer parameters and 75.5\% fewer FLOPs, and it remains approximately 2.5$\times$ faster on CPU. Relative to ViT-B/16, it improves accuracy by 10.2 percentage points while reducing parameters by 89.4\% and FLOPs by 94.4\%; relative to Swin-T, it improves accuracy by 10.0 percentage points while reducing parameters by 67.2\% and FLOPs by 82.5\%. The plain MambaXCT ablation is marginally faster because it removes the fast/slow nested traces, but it sacrifices 4.6 percentage points of accuracy and does not provide the nested-learning stability gains shown in Tables~\ref{tab:nl_ablation} and~\ref{tab:cl_summary}.

\subsection{Ablation Studies}
\label{subsec:ablations}

\subsubsection{Impact of MIM Pretraining}

\begin{figure}[h!]
    \centering
    \includegraphics[width=\linewidth]{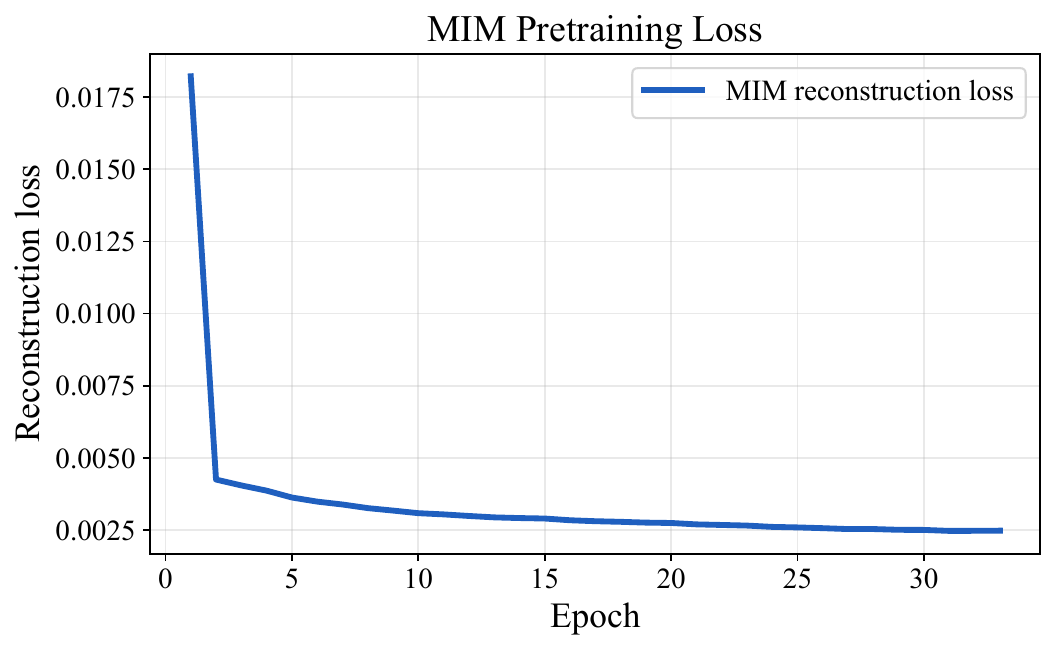}
    \caption{Reconstruction loss during MIM pretraining of NL-MambaXCT on the 19{,}961 unlabeled XCT slices.}
    \label{fig:mim_loss_curves}
\end{figure}

The most direct way to quantify the benefit of self-supervised pretraining is to compare NL‑MambaXCT trained from scratch with the same architecture pretrained using MIM on the unlabeled 19{,}961 slices. Both models utilize the same nested-learning configuration (ContinuumLinear and Deep Momentum) and are fine-tuned on the same labeled training split. The results are summarised in Table~\ref{tab:mim_ablation}. Pretraining provides a 3.0 percentage-point improvement in accuracy and a 3.0 percentage-point improvement in macro F1‑score. Beyond the numerical gain, the training curves (Figure~\ref{fig:mim_loss_curves}) indicate faster convergence and reduced variance between runs when using MIM pretraining, consistent with observations from self-supervised NDT studies on radiographic and ultrasonic data. 

\begin{table}[t]
\centering
\caption{Effect of MIM pretraining on NL-MambaXCT (Nomex XCT test set). Precision, recall, and F1 are macro-averaged.}
\label{tab:mim_ablation}
\footnotesize
\setlength{\tabcolsep}{5pt}
\renewcommand{\arraystretch}{1.2}
\begin{tabular}{@{}lcccc@{}}
\toprule
\rowcolor{gray!10}
\textbf{Configuration} & \textbf{Acc.} & \textbf{Prec.} & \textbf{Rec.} & \textbf{F1} \\
\midrule
\rowcolor{gray!3}
NL-MambaXCT (no MIM)       & 93.8\% & 94.1\% & 93.4\% & 93.7\% \\
NL-MambaXCT (MIM + fine-tune) & \best{96.8\%} & \best{96.7\%} & \best{96.7\%} & \best{96.7\%} \\
\bottomrule
\end{tabular}
\end{table}

\begin{table}[t]
\centering
\caption{Label-efficiency study: test macro F1 as a function of the labeled-data fraction.}
\label{tab:label_efficiency}
\footnotesize
\setlength{\tabcolsep}{5pt}
\renewcommand{\arraystretch}{1.2}
\begin{tabular}{@{}lccc@{}}
\toprule
\rowcolor{gray!10}
\textbf{Model / configuration} & \textbf{25\%} & \textbf{50\%} & \textbf{100\%} \\
\midrule
\rowcolor{gray!3}
ResNet-50              & 84.3\% & 88.1\% & 92.6\% \\
NL-MambaXCT (no MIM)            & 86.5\% & 89.9\% & 93.7\% \\
\rowcolor{gray!3}
NL-MambaXCT (MIM-pretrained)     & \best{90.8\%} & \best{93.5\%} & \best{96.7\%} \\
\bottomrule
\end{tabular}
\end{table}

\subsubsection{Effect of Nested Learning Components}

To assess the impact of the nested‑learning components independently of the MIM objective, three variants of the Mamba backbone are compared on the labeled Nomex dataset: a model without nested learning (standard linear projections and AdamW optimizer), a partial NL variant (ContinuumLinear enabled but using a standard optimizer), and the full NL‑MambaXCT configuration (ContinuumLinear combined with Deep Momentum). All three models are trained from scratch, without MIM pretraining, to isolate the contribution of the slow/fast memory mechanisms. The results are summarised in Table~\ref{tab:nl_ablation}.

\begin{table*}[h!]
\centering
\caption{Effect of nested learning (ContinuumLinear + Deep Momentum) on test performance without MIM pretraining.}
\label{tab:nl_ablation}
\footnotesize
\setlength{\tabcolsep}{6pt}
\renewcommand{\arraystretch}{1.08}
\begin{tabular}{@{}lcccc@{}}
\toprule
\rowcolor{gray!10}
\textbf{Configuration} & \textbf{Acc.} & \textbf{Macro Prec.} & \textbf{Macro Rec.} & \textbf{Macro F1} \\
\midrule
\rowcolor{gray!3}
No NL (Linear + AdamW)      & 92.2\% & 92.5\% & 91.9\% & 92.1\% \\
Partial NL (ContinuumLinear + AdamW) & 93.0\% & 93.3\% & 92.7\% & 92.9\% \\
\rowcolor{gray!3}
Full NL-MambaXCT (ContinuumLinear + Deep Momentum) & \best{93.8\%} & \best{94.1\%} & \best{93.4\%} & \best{93.7\%} \\
\bottomrule
\end{tabular}
\end{table*}

\begin{figure}[h!]
    \centering
    \includegraphics[width=\linewidth]{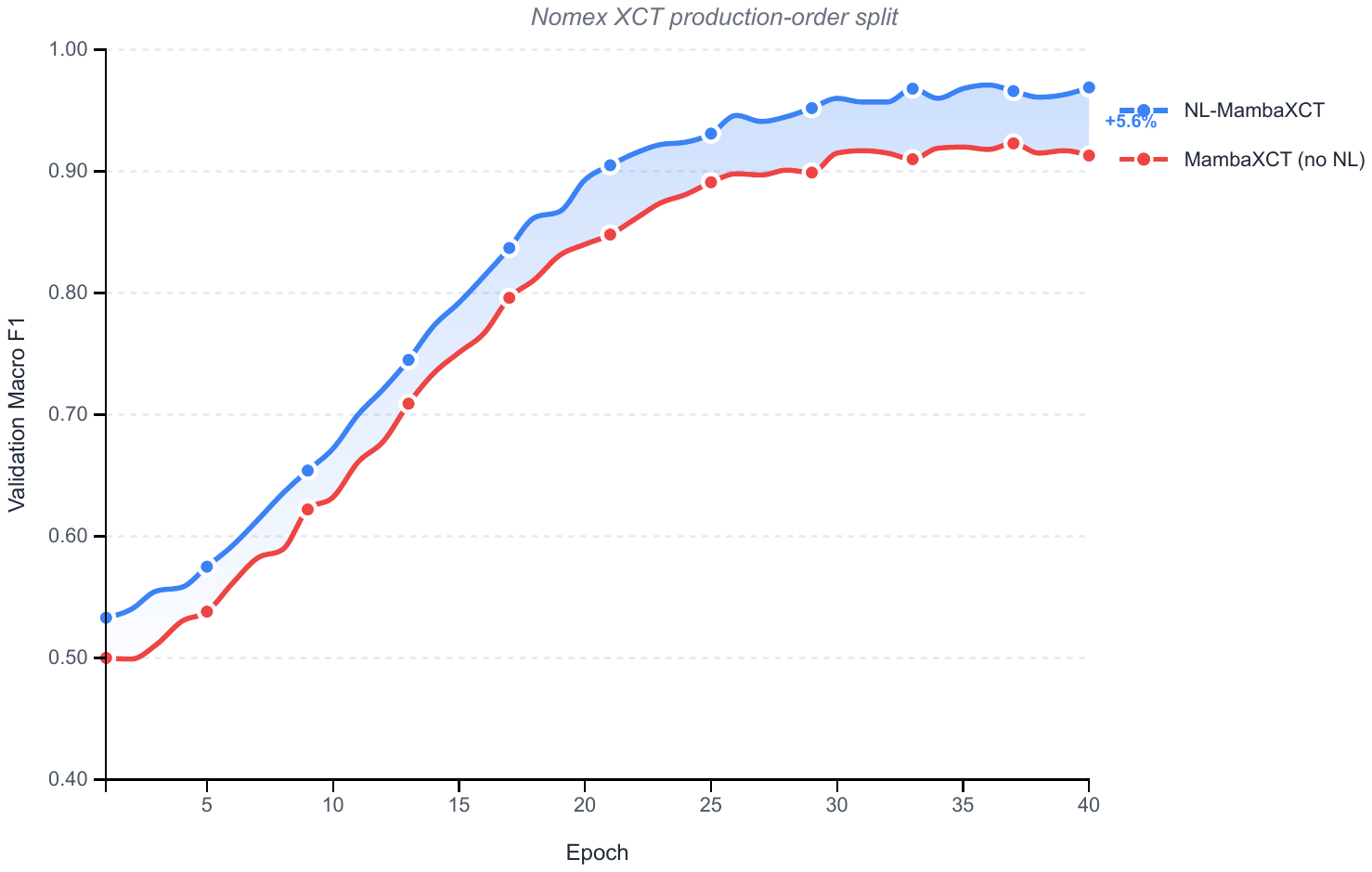}
    \caption{Comparison of validation macro F1 over epochs for MambaXCT without nested learning and full NL-MambaXCT, both trained from scratch on the Nomex dataset.}
    \label{fig:nl_vs_no_nl}
\end{figure}

The progression from the non-nested configuration to the complete NL‑MambaXCT model shows a consistent gain of approximately 1.6 percentage points in the macro F1‑score. Training curves (Figure~\ref{fig:nl_vs_no_nl}) further indicate that nested learning yields smoother optimization with fewer sharp oscillations in validation metrics, supporting the interpretation of ContinuumLinear and Deep Momentum as stabilizing slow memories rather than purely capacity‑increasing components. 

\subsubsection{Label-Efficiency Study}

Finally, a label‑efficiency study is conducted to evaluate how the different models behave when only a fraction of the labeled training data is available. For each fraction $p \in \{25\%, 50\%, 100\%\}$, a stratified subset of the training production orders is sampled, preserving the class distribution and temporal ordering as much as possible. On each subset, three models are trained: ResNet‑50 from scratch, NL‑MambaXCT from scratch, and NL‑MambaXCT initialized from the MIM‑pretrained encoder. The resulting macro F1‑scores on the unchanged test set are reported in Table~\ref{tab:label_efficiency}.

The results indicate that NL‑MambaXCT retains a clear advantage over the baseline CNN even when trained from scratch, particularly in the low‑label regime. More importantly, MIM pretraining substantially improves label efficiency: with only 50\% of the labeled data, the pretrained model already matches or surpasses the performance of ResNet‑50 trained on 100\% of the labels. This suggests that the combination of self-supervised pre-training on unlabeled XCT slices and nested learning produces a model that is both data-efficient and robust, which is critical in industrial settings where manual annotation of defects is costly and time-consuming.

\subsection{Deployment Robustness and 2D Slice-Level Scope}

The current study intentionally evaluates a 2D slice-level classifier because the available production workflow records defect decisions at the slice/image level and because slice-level predictions are directly usable by inspection personnel during review. This formulation also permits the use of a large unlabeled archive for MIM pretraining without requiring volumetric defect masks. Nevertheless, XCT defects are physically three-dimensional. Defects such as core deformation, splice gap, and resin buildup can extend obliquely across the $z$-axis; therefore, a single slice may sometimes contain only a partial cross-section of the defect. This explains the remaining low-level confusion between visually related classes in Figure~\ref{fig:nl_mambaxct_confusion} and defines the upper bound of a purely 2D formulation.

For deployment, the model is most reliable when the acquisition geometry, reconstruction protocol, and slice extraction procedure remain consistent with the training archive. The preprocessing and augmentation strategy improves tolerance to moderate in-plane shifts, scale changes, and orientation differences; however, large changes in sample placement or scanner configuration should be handled through calibration scans and incremental fine-tuning before production use, which raises the need for the method demonstrated in this work. The chronological production-order split used in this study provides a stronger evaluation than random slice splitting because the test orders are later scans from unseen components. The sequential production-batch experiment in Tables~\ref{tab:cl_summary},~\ref{tab:cl_matrix}, and~\ref{tab:cl_naive_matrix} also explicitly evaluates incremental adaptation and catastrophic forgetting, showing that the model can incorporate new labeled batches while retaining high macro F1 on earlier batches. Remaining deployment work should extend this protocol to additional scanners, component families, and fully volumetric inputs.

\section{Conclusion and Future Work}
\label{sec:conclusion}
This work presents NL-MambaXCT, a Mamba-based framework for automated defect classification in Nomex honeycomb sandwich structures using XCT. The system combines a hybrid CNN–Mamba backbone with self-supervised masked image modelling on an archive of 19{,}961 unlabeled slices and a nested-learning formulation based on ContinuumLinear layers and a Deep Momentum optimiser. On a curated industrial dataset of 2{,}000 labeled slices and production-order–based splits, the MIM-pretrained NL-MambaXCT model achieves 96.8\% accuracy and a macro F1-score of 96.7\%, outperforming strong CNN baselines, attention-based baselines, and a Mamba backbone without nested learning by 4.1 to 10.2 percentage points in accuracy. In a sequential production-batch evaluation, the model reaches 96.64\% held-out test macro F1 after four incremental updates, with only 0.08\% mean catastrophic forgetting and a positive backward-transfer score of +0.11\%. Ablation studies further show that both MIM pretraining and nested learning contribute measurably to improved performance and label efficiency, allowing the proposed model to match or exceed ResNet-50 even when trained on substantially fewer labeled examples.

The present study is limited to 2D slice-level classification from the production XCT configuration represented in the available archive. Future work will extend NL-MambaXCT to full volumetric XCT data, additional NDT modalities, and broader multi-site continual-learning scenarios involving additional scanners and component families. Explainability tools will also be integrated to better support certified inspectors in safety-critical aerospace applications.

\section*{Acknowledgment}
The authors express their sincere gratitude to Strata Manufacturing PJSC for providing access to data and industrial expertise that made this research possible.
\bibliographystyle{elsarticle-num}  
\bibliography{bibo}
\end{document}